\documentclass[final,authoryear,times,twocolumn]{elsarticle}

\usepackage{amssymb,amsmath,graphicx}

\journal{Journal of Air Transport Management}

\begin{document}

\begin{frontmatter}



\title{Experimental test of airplane boarding methods}


\author{Jason H. Steffen}
\address{Fermilab Center for Particle Astrophysics, Batavia, IL}
\ead{jsteffen@fnal.gov}

\author{Jon Hotchkiss}
\address{Hotchkiss Industries, Sherman Oaks, CA}
\ead{jon@jonhotchkiss.com}

\begin{abstract}
We report the results of an experimental comparison of different airplane boarding methods.  This test was conducted in a mock 757 fuselage, located on a Southern California soundstage, with 12 rows of six seats and a single aisle.  Five methods were tested using 72 passengers of various ages.  We found a significant reduction in the boarding times of optimized methods over traditional methods.  These improved methods, if properly implemented, could result in a significant savings to airline companies.
\end{abstract}

\begin{keyword}
Airplane Boarding



\end{keyword}

\end{frontmatter}


\section{Introduction}
\label{introduction}

The process of boarding an airplane is of interest to a variety of groups.  The public is interested both as a curiosity, as it is something that they may regularly experience, and as a consumer, as their experiences good or bad can affect their loyalties.  Airline companies and their employees also have a stake in an efficient boarding procedure as time saved in the boarding process may result is monetary savings, in the quality of interactions with passengers, and in the application of human resources to the general process of preparing an airplane for departure.

A recent study \citep{Nyquist:2008} indicates that the average cost to an airline company for each minute of time spent at the terminal is roughly \$30.  Thus, each minute saved in the turn-around time of a flight has the potential to generate over \$16,000,000 in annual savings (assuming an average of 1500 flights per day).  While the boarding process may not be the primary source of delay in returning an airplane to the skies, reducing the boarding time may effectively eliminate passenger boarding as a contributor in any meaningful measure.  Consequently, subsequent efforts to streamline the other necessary tasks, such as refueling and maintenance, would be rewarded with a material reduction in time at the gate for each flight.

Several studies of the airplane boarding process exist \citep{Nagel:2005,VandenBriel:2005,Bazargan:2007,VanLandeghem:2002,Steffen:2008a,Steffen:2008b,Bachmat:2006} and many of the conclusions are universal.  The optimization is essentially a reduction of the number of times that passengers must either wait for or traverse each other, whether in the aisle (an aisle interference) or within a given row of seats (a seat interference).  Some methods resulting from these studies are having the passengers seated at the windows boarding first, followed by the middle and aisle seats (hereafter called ``Wilma'').  Another method is the ``Reverse Pyramid'' method which adds an emphasis on boarding the rear of the cabin first \citep{VandenBriel:2005}.  Both the Wilma method and the Reverse Pyramid method specifically eliminate seat interferences and, to differing degrees, aisle interferences.

Many of these previous studies concentrated on methods involving boarding groups rather than having passengers line up in a specified order.  Thus, the authors could identify the best method of those that were tested, but it was not known if those methods represented the optimum boarding method overall.  \citet{Steffen:2008a} claims to have identified the optimum boarding method under certain assumptions.  This method (which we call the Steffen method) added to the reduction of aisle and seat interferences the idea of efficient, parallel use of the aisle.  For example, if passengers are ordered such that the plane boards from the rear window seats, row-by-row to the front aisle seat there would be no seat interferences and no passenger would need to pass another in the aisle.  However, in this scenario only the lead passenger or two would be able to stow their luggage---the rest of the passengers would simply be filling the aisle (rather than aisle interferences being eliminated, they are universal).

The Steffen method, on the other hand, orders the passengers in such a way that adjacent passengers in line are sitting in corresponding seats two rows apart from each other (e.g., 12A, 10A, 8A, 6A, etc.).  This method trades a small number of aisle interferences at the front of the cabin, for the benefit of having multiple passengers stowing their luggage simultaneously.  Other methods, such as Wilma and the Reverse Pyramid also realize parallel use of the aisle in a natural way as adjacent passengers are frequently sitting in widely separated rows.

Practical implementation aside, the Steffen method claims to be the fastest possible method, on average, to board passengers onto an airplane with a single door and a single aisle.  We test that claim here.  Other claims made in \citet{Steffen:2008a} that we test in this experiment include: 1) random boarding (where passengers have assigned seats, but are allowed to board at any time) should perform as well as Wilma and 2) that boarding in blocks should perform worse than Wilma, random, and the Steffen method.  Both of these secondary predictions are based upon efficiency considerations in the use of the aisle to stow luggage---random, Wilma, and the Steffen method spread passengers throughout the cabin while boarding in blocks tends to concentrate passengers in one part of the aircraft.  We note that \citet{Steffen:2008a} does not fully treat seat interferences---which will be a likely source of discrepancy between the predictions in \citet{Steffen:2008a} and reality.  Another claim that we were not able to test, but which could be part of a future study, is that due to inefficiency in the use of the aisle, boarding from the back to the front of the cabin is nearly as bad as boarding from the front to the back.

This article is organized as follows.  In section \ref{setup} we describe the facility we used to conduct the test, the geometry of the mock airplane, and the assignments given to the passengers.  Section \ref{methods} discusses details of the different methods we tested and how they were implemented.  We present the results of our test in section \ref{results} and discuss them in some detail in section \ref{discussion}.  Our concluding remarks are given in section \ref{conclusions}.

\section{Experimental Setup}
\label{setup}

We conducted this experiment at the \textit{Air Hollywood} soundstage in Studio City, CA.  The narrow-body mock airplane was used which had 12 rows of six seats with a single, central aisle.  A small first-class section (two rows of seats) was used for camera and lighting equipment.  The width of the fuselage is 11' 7''.  Each seat has the standard 17'' width not including the 2'' armrest.  Rows of seats are spaced 32'' and the aisle was 21'' wide.  The overhead bins were standard for older-model aircraft, but are slighly smaller than modern bins.  In particular, the overhead bins did not accomodate standard size roll-aboard luggage when inserted wheels first.

We employed 72 passengers---volunteers and hollywood extras.    The ages of the passengers ranged from young children age five through retired seniors with the bulk of passengers being employment-age adults (we did not solicit or record passenger ages).  Passengers generally had a bag, roll-aboard carry-on, or both---though a small number of passengers had no luggage.  When all passengers were on board with their luggage stowed there was very little remaining space in the overhead bins---essentially none of which was useful for additional storage.  Thus, when the airplane was full, so were the overhead bins; though, there was never a time when a piece of luggage would not fit somewhere in the aircraft.  Since excess luggage would likely affect each boarding method in a similar way the relative ranking of the different methods---in terms of the boarding time---should be well established by our test.

Each passenger was given a set of five tickets with a seat assignment and either a passenger number or boarding group number (depending upon the method being tested).  The methods were labelled numerically so that the passengers did not know which method was being tested.  Seat assignments were chosen randomly such that passengers did not sit in the same seat with each method nor were they deliberately placed in the same location in the line or in the same boarding group.  The only exceptions to these rules were three parent-child pairs who always had two adjacent seats and who were always at the front of the line, boarding first.

Finally, passengers used their individual luggage for each method (they did not swap luggage with other passengers).  One could argue that swapping luggage would make the test of each method more robust.  However, it is not clear that this is the case as fatigue with boarding the aircraft several times could mitigate the benefit of practice.  Moreover, the adopted scenario is more representative of actual airplane boarding as passengers are likely to have the same or similar bags with them whenever they travel.  Also, since passengers were generally not sitting in the same seats and did not board at the same time with each method, the luggage that was already stowed in the overhead bins would be different with each test---adding, to some extent, the desired randomness.

\section{Boarding Methods}
\label{methods}

\begin{figure}
\begin{center}
\includegraphics{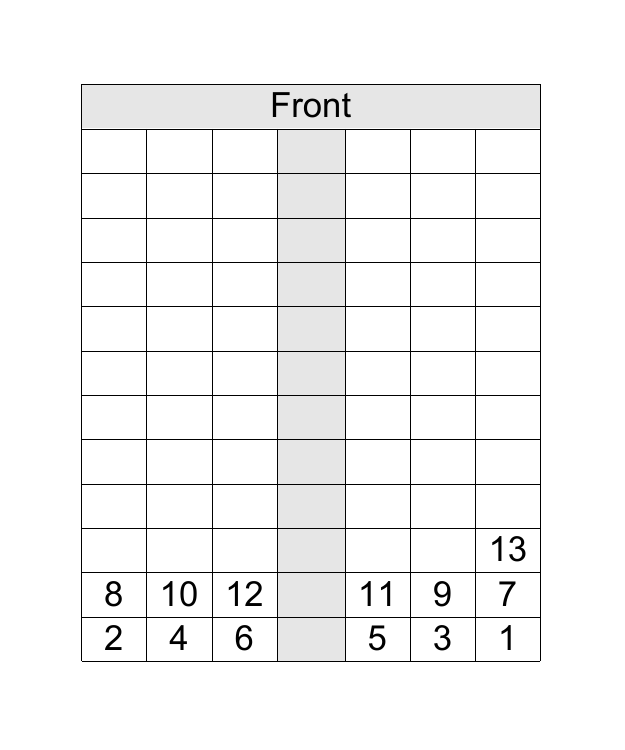}
\caption{Back to front seating order.\label{method1}}
\end{center}
\end{figure}

\begin{figure}
\begin{center}
\includegraphics{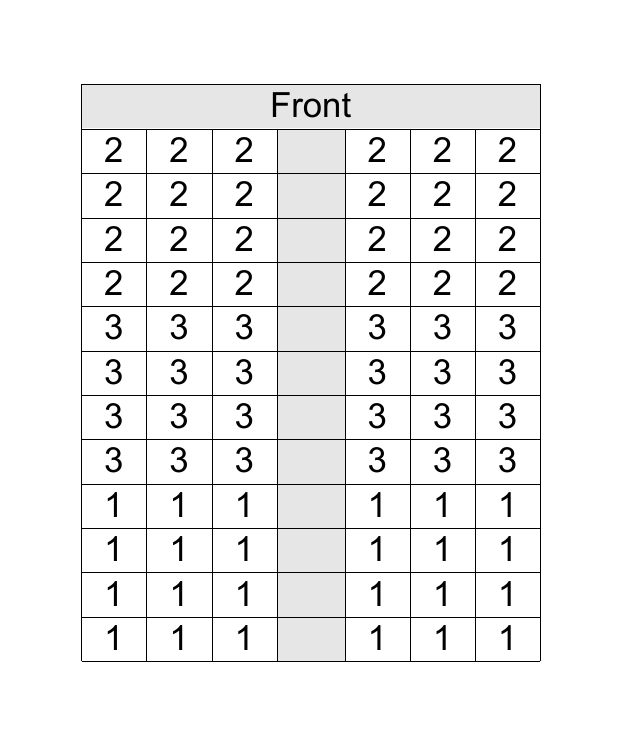}
\caption{Block boarding groups.\label{method2}}
\end{center}
\end{figure}

\begin{figure}
\begin{center}
\includegraphics{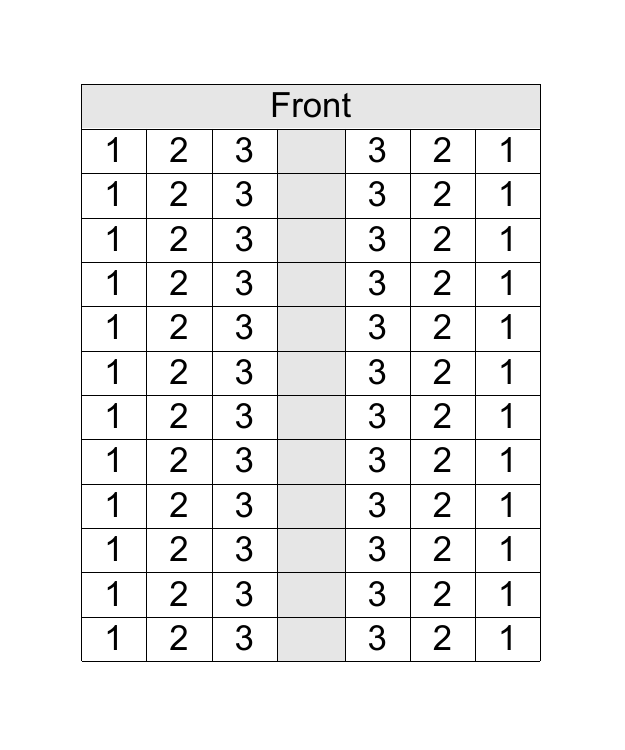}
\caption{Wilma boarding groups.\label{method3}}
\end{center}
\end{figure}

\begin{figure}
\begin{center}
\includegraphics{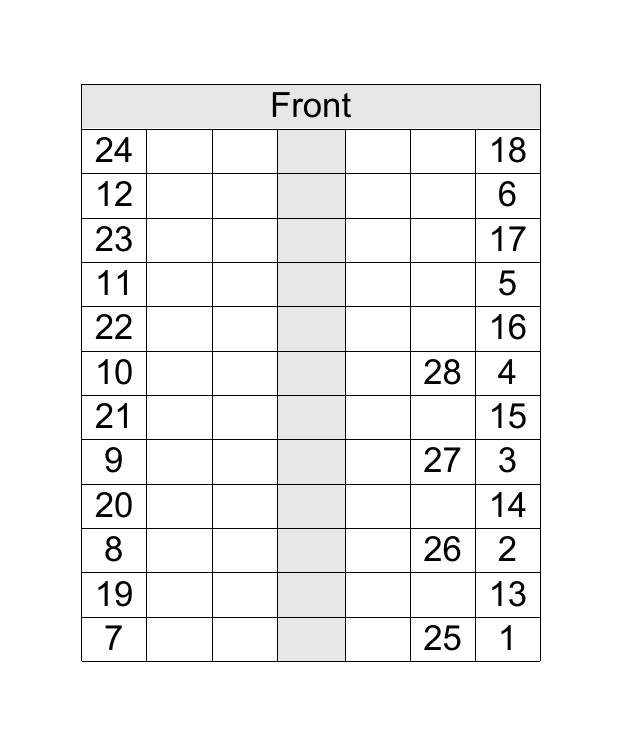}
\caption{Steffen method seating order.\label{method4}}
\end{center}
\end{figure}

\begin{figure}
\begin{center}
\includegraphics{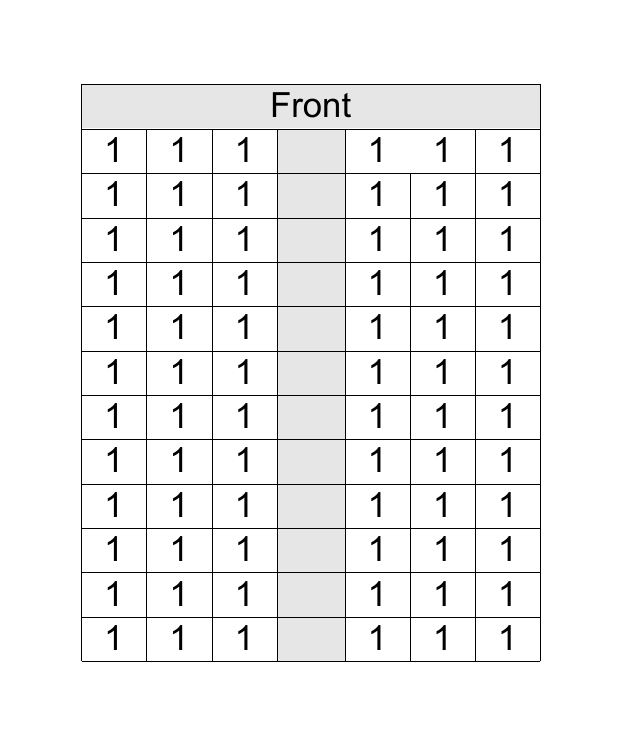}
\caption{Random method boarding group.\label{method5}}
\end{center}
\end{figure}

The five methods that were tested include: 1) boarding from the back to the front of the aircraft in a specified order, 2) boarding in four-row blocks, 3) the Wilma method, 4) the Steffen method, and 5) random boarding.  Each of these methods is described below.  Figures \ref{method1}, \ref{method2}, \ref{method3}, \ref{method4} and \ref{method5} show the seating assignments or boarding groups.

The back-to-front method had passengers in a specific order starting with the back row window seats, then middle seats, then aisle seats.  This pattern repeated itself for the rest of the rows moving toward the front of the plane.  A naive guess regarding the most efficient method to board an airplane might suggest that this method is optimal as it is orderly, boarding passengers from the back of the plane to the front and from the windows to the aisle, and it avoids all seat interferences---excepting priority boarding.

The block method divided the 12 rows into three groups of four rows.  The back four rows were the first group, followed by the front four rows, and finishing with the center four rows.  The decision to use this method rather than boarding in blocks from the back to the front was admittedly arbitrary, but we believe that the results should be representative of either method.  We also elected to use three boarding groups rather than four so that each group would involve a larger number of rows---making it more similar to what one would expect with a full-size aircraft and capturing more the potential benefit of having more rows available to the passengers in each group.

The Wilma method used three boarding groups with the first group being all window seats, the second being all middle seats, and the third being all aisle seats.  Within each group the passengers were essentially random.  Thus, this method when followed would avoid passenger seat interferences but would have passenger aisle interferences.

The Steffen method has the passengers lining up in a prescribed order.  This order incorporates in a specific way boarding from the back to the front and from the windows to the aisle.  Adjacent passengers in line are sitting two rows apart from each other in corresponding seats (e.g. 12F, 10F, 8F, 6F, 4F, 2F).  This method attepts to eliminate seat interferences and, as much as possible, aisle interferences while naturally allowing multiple passengers to stow their luggage simultaneously.  The separation between adjacent passengers provides some space for each passenger to manipulate their luggage into the bins.

The final method was random boarding with a single boarding group.  Since each passenger still had an assigned seat, this method is not equivalent to the free-for-all method where passengers select their seats at will.  There is an important distinction between random boarding (where the passengers have assigned seats but not an assigned place in line) and free-for-all boarding.  Free-for-all boarding is not random since passengers can make decisions about where to sit once they observe the state of the cabin---passengers can choose to avoid an aisle or seat interference where passengers in the random boarding method cannot.

\section{Results}
\label{results}

For each method there are two reported boarding times.  The first ''official'' time is the elapsed time from when the first passenger set foot in the aircraft until the last passenger was seated.  The second ''extended'' time is the elapsed time from when the passengers were told to proceed to board until the last passenger was seated.  These times differ by only a few seconds and we report both times here; tnhe difference representing the time it took passengers to walk approximately six feet and to climb two stairs into the airplane.  Table \ref{times} shows the results of the different boarding methods.

\begin{table}
\begin{center}
\caption{Elapsed time for each boarding method in minutes and seconds.}\label{times}
\begin{tabular}{ccc}\hline
Method & Official & Extended \\
 & Time & Time \\ \hline \hline
Back--Front   & 6:11 & 6:16 \\
Blocks        & 6:54 & 6:56 \\
Wilma         & 4:13 & 4:21 \\
Steffen       & 3:36 & 3:40 \\
Random        & 4:44 & 4:48 \\ \hline
\end{tabular}
\end{center}
\end{table}

We recognize that there are random fluctuations in the boarding process and that, in the absence of multiple realizations of the experiment our boarding times are only estimates of the mean boarding time for each method.  It was not feasible to re-run each boarding method several times.  As the variance in the boarding time comes from a series of random interferences, a reasonable estimate of the fractional uncertainty in our estimate of the mean boarding time is approximately $1/\sqrt{72} \sim 10\%$.

We note that there may be some systematic biases at play in our results (for example, the quantity and size of the luggage that our passengers carried may be somewhat different than what a similar group might take while traveling).  This difference, and other differences are more likely to affect the total length of time each method takes to board than they are to affect the ratios of those times.  Thus, while our times may be systematically low or high compared to what one would experience at a real airport with real travelers, the ratios of those times should be very similar in the different environments.  Regardless, given the full capacity of both the airplane and the overhead storage areas, the wide range in passenger ages, and the random nature of the experiment, we believe that these potential sources of bias will have a relatively small effect on the boarding times and that our results for each method are sufficient for both qualitative discussion and for rigorous comparisons among them.

\section{Discussion}
\label{discussion}

The boarding time for back-to-front boarding is an estimate of the effect of aisle interferences---which is, in turn, an estimate of the average luggage loading time for our sample of passengers.  This equivalency between luggage loading time and the time of an aisle interference is only applicable in this method because here the aisle interference happens at or near the row of the blocked passenger (we discuss this more below).  Nominally, back-to-front boarding has no seat interferences, but it does have 71 aisle interferences.  This method took just over 6 minutes (360 seconds---allowing the balance of the measured time to account for the time required to walk the length of the aircraft) which gives roughly 5 seconds per interference.  Therefore, an estimate of the effective time for a passenger to stow luggage upon arriving at their seat is also 5 seconds---the actual time is likely slightly longer as there arise situations where multiple passengers stow their luggage simultaneously.

The difference in boarding time between random boarding and Wilma gives insight into the delay caused by seat interferences.  Wilma nominally has no seat interferences while random boarding does have some.  There are $3! = 6$ different permutations of the order that passengers sit in their half-row (the three seats on their side of the aisle).  The number of seat interferences caused by each permutation ranges from 0 to 3 with a mean of 1.5 (we will assume for simplicity that the delays for seat interferences in a particular half-row add linearly rather than, say, in quadrature).  Thus, a typical passenger in random boarding will be involved in 1.5 seat interferences.  However, in relation to the time required to fill the aircraft, not all seat interferences are equivalent.  Only seat interferences that cause an aisle interference will have a meaningful affect on the boarding time---a seat interference at the back of the airplane will not block a passenger who is stowing luggage at the front.  Approximately half of the seat interferences will cause such an aisle interference (the probability that the subsequent passenger has a seat in a row number greater than or equal to the blocking passenger is actually $1/2 + 1/N$, where $N$ is the number of rows in the aircraft---this subtlety will be important later, but here we are simply trying to get a rough estimate of the cost of a seat interference).

Between Wilma and random boarding we would expect to have $1/2 \times 1.5 \times 72 = 54$ meaningful seat interferences.  The observed difference in boarding times was roughly 30 seconds such that the effective delay for a seat interference, in this case, is just under 1 second.  The true delay is actually twice this quantity since only about half of the interferences affect the boarding time.  At first glance this would seem an underestimate for the time it takes for a seated passenger to stand and switch places with a second passenger.  However, much of this maneuvering is done in parallel with the luggage stowing---a passenger typically informs their neighbor, before their luggage is completely stowed, that they will need to move past one another.  Thus, the two-second delay is primarily for passengers to file back into their half-row---a reasonable estimate.

Taking these effects into consideration, one may wonder why block boarding fared so poorly.  Block boarding allows for some passengers to stow their luggage simultaneously---eliminating some aisle interferences.  The particular method we employed also eliminates some aisle interferences by having the first and second groups sit in different portions of the aircraft.  Here is where the subtle difference between random boarding and block boarding might provide an explanation.  In random boarding the length of the airplane was 12 rows.  In block boarding you are essentially boarding three smaller airplanes of only four rows.  Thus, the probability for a particular seat interference to cause an aisle interference grows from $1/2+1/12 = 0.58$ to $1/2 + 1/4 = 0.75$ which is nearly 30\% higher.  The number of passengers who can simultaneously stow their luggage is at most two, if one assumes that a passenger needs two rows of space to do so.  In random boarding this number grows to six simultaneous passengers---again at most (typically these numbers will be somewhat less).

Thus, if the number of rows in a boarding block is too small, the result is little different than boarding back to front.  But, now there are a considerable number of seat interferences, roughly 30\% more for our airplane geometry.  The extreme case is block boarding by single rows where the aisle interferences are exacly what one would get from back to front, but now one expects three seat interferences at the full $\sim 2$ second cost of a seat interference (all seat interfereces affect all passengers).  While there is likely some random element to the nearly 45 second increase in boarding time from back to front compared with block boarding, the effect of seat interferences alone would account for nearly 40 of those seconds (30\% more than the 30 second difference observed between random and Wilma).  Clearly, boarding in blocks of four rows does not help the enplaning process.  Blocks of 12 rows, on the other hand, clearly does---the difference between back to front and random boarding (almost 90 seconds) shows this.  Said another way, boarding blocks of 12 rows (our entire airplane) instead of blocks of four rows eliminates the equivalent of 18 passengers stowing their luggage (due to a combination of reducing the number of aisle interferences and adding a number of seat interferences).  Still, in terms of fast boarding times, block boarding will likely never be superior to random boarding---only other considerations would serve to justify its use.

How does the Steffen method compare?  In our test, this method was considerably better than all other methods.  By design the Steffen method has no seat interferences and 11 aisle interferences---passengers enter in 12 sets, but the first set has no interference preceeding it.  Yet, even in the controlled environment of our experiment, this design was not fully implemented.  Six passengers (those with small children) boarded first, some passengers will sit in the wrong seat.  In short, there is still randomness to the process.  This randomness alone cannot explain why the Steffen method did not board in a single minute as one might estimate from eleven aisle interferences of five seconds each.

The primary answer lies in a remaining, characteristic timescale in the boarding process---the time it takes to walk the length of the airplane.  This last part is manifest in the nature of the aisle interferences.  For back-to-front boarding, the aisle interference corresponds to a 5 second wait for the preceeding passenger, after which the blocked passenger can immediately begin stowing their luggage.  For random boarding a typical aisle interference corresponds to the same 5 second wait, but now it is, on average, in the middle of the aircraft.  This wait is followed by a walk of roughly half of the remaining distance to the rear of the cabin to the average location of the passenger's row.  The aisle interferences in the back to front method are less severe as there is no subsequent walk.  In the Steffen method the aisle interferences are of the worst sort.  They occur at the front of the airplane with the entire length of the airplane to traverse once the interference has passed.

If one assumes that it takes roughly one second to walk past each row, then the Steffen method has nearly 2.5 minutes of built-in delays just from walking.  Add the luggage loading time of five seconds at the start of each delay and you are beyond three minutes.  Add the seat interferences from the parent/child pairs and an estimate consistent with the observed boarding time of roughly 3.5 minutes takes shape.

From the Steffen method to the Wilma method one trades an increase in the number of aisle interferences for aisle interferences that are less severe (the result of this trade, however, still increases the boarding time).  Then, from Wilma to random boarding one adds seat interferences (for the benefit of simplicity and perhaps some unknown amount of customer satisfaction).  Block boarding, finally, adds cost to the seat interferences, making them relatively worse in terms of the time they consume as a larger fraction of the seat interferences will cause aisle interferences.

We note that the estimates we have derived for the time consumed by walking the aisle, seat and aisle interferences, and luggage stowing are all estimates based upon the observed boarding times of the methods.  They were made in an effort to properly interpret the results in terms of known factors in the boarding process.  A more rigorous study of each of these quantities is beyond the scope of this work, but would serve to provide more understanding the interplay between all of the factors, including some not discussed here, that contribute to the boarding time.

\section{Conclusions}
\label{conclusions}

We have seen experimentally that there is a marked difference in the time required to board an aircraft depending upon the boarding method used.  The evidence strongly supports the heuristic argument from \citet{Steffen:2008a} that methods that parallelize the boarding process by more efficiently utilizing the aisle (having more passengers stow their luggage simultaneously) will board more quickly than those that do not.  The relative benefit of the application of this theory will grow with the length of the aircraft.  Here, we used a 12-row mock airplane, but a more typical airplane with twice that number of rows will gain more by the implementation of parallelized boarding methods.

How this improvement scales with the cabin length is different for each method.  For the Steffen method, the benefit will scale almost linearly.  If the airplane is twice as long, the time savings will be nearly twice as much since the density of luggage-stowing passengers will remain the same and the boarding will still be maximally parallel.  For Wilma and random boarding the benefit will not be as strong since the benefits of parallel boarding are randomly distributed along the length of the cabin instead of being regularly distributed.  For block boarding the benefit will come almost exclusively through an increased size of the boarding group (more rows per group).

Given the observed boarding times, the Wilma method boards faster than block boarding (with four-row blocks) by a factor of almost 1.7 while the Steffen method boards faster by almost a factor of 2.  An accurate estimate of the time it would take to board an airplane two or three times as long would take a more detailed study of the relative interplay of the relevant timescales of the problem---the time to stow luggage (i.e., the time consumed by an aisle interference), the time consumed by seat interferences, and the time required to walk the length of the airplane.  Nevertheless, even if one simply doubles the boarding times observed here the savings of the Steffen method over the block method would be nearly seven minutes.  This savings could be as much as \$110,000,000 annually per carrier---well over a billion dollars for the industry---and likely could be more given the parallel nature of the boarding process.  Indeed, a test with a longer aircraft may show surprising results in this regard.

Some practical impedements arise when implementing the Steffen method.  However, it was fairly easy to do in our experiment, and some carriers already have the customers line up in a specified order (e.g., Southwest Airlines and, more recently, a trial by American) showing that the challenge is surmountable.  Other considerations such as companion travelers are also fairly simple to accomodate in the context of the Steffen method and, if done well, would have a marginal, if not negligible, effect on the boarding time.  (Specifically, as long as paired passengers are not seated near the front of the aircraft where the primary aisle interferences materialize, they will have a few additional seconds---on average half of the time required to walk the length of the cabin---to stow their luggage and be seated.)  While the cost savings of the Steffen, or other optimized boarding method, may not be completely realizable due to maintenance and outfitting time, the motivation to reduce the time required for these other, necessary activities would be increased if passenger boarding was removed as an effective limitation.

\section*{Acknowledgements}

We wish to thank the staff at Hotchkiss Industries for their work in preparing and executing this experiment.  We also wish to thank the many participants who devoted the greater part of a day to the project---their absence would have been severely missed.

\bibliographystyle{elsarticle-harv}
\bibliography{airplaneboarding}

\begin{thebibliography}{8}
\expandafter\ifx\csname natexlab\endcsname\relax\def\natexlab#1{#1}\fi
\expandafter\ifx\csname url\endcsname\relax
  \def\url#1{\texttt{#1}}\fi
\expandafter\ifx\csname urlprefix\endcsname\relax\def\urlprefix{URL }\fi

\bibitem[{{Bachmat} et~al.(2006){Bachmat}, {Berend}, {Sapir}, {Skiena}, and
  {Stolyarov}}]{Bachmat:2006}
{Bachmat}, E., {Berend}, D., {Sapir}, L., {Skiena}, S., {Stolyarov}, N., Jul.
  2006. {Analysis of aeroplane boarding via spacetime geometry and random
  matrix theory}. Journal of Physics A Mathematical General 39, L453--L459.

\bibitem[{Bazargan(2007)}]{Bazargan:2007}
Bazargan, M., 2007. A linear programming approach for aircraft boarding
  strategy. European Journal of Operational Research 183~(1), 394--411.

\bibitem[{Landeghem and Beuselinck(2002)}]{VanLandeghem:2002}
Landeghem, H.~V., Beuselinck, A., 2002. Reducing passenger boarding time in
  airplanes: A simulation based approach. European Journal of Operational
  Research 142~(2), 294 -- 308.

\bibitem[{Nagel and Ferrari(2005)}]{Nagel:2005}
Nagel, K., Ferrari, P., 2005. The secrets to faster boarding. International
  Airport Review 3.

\bibitem[{Nyquist and McFadden(2008)}]{Nyquist:2008}
Nyquist, D.~C., McFadden, K.~L., 2008. A study of the airline boarding problem.
  Journal of Air Transport Management 14~(4), 197 -- 204.

\bibitem[{Steffen(2008a)}]{Steffen:2008a}
Steffen, J.~H., 2008a. Optimal boarding method for airline passengers. Journal
  of Air Transport Management 14~(3), 146 -- 150.

\bibitem[{{Steffen}(2008b)}]{Steffen:2008b}
{Steffen}, J.~H., 2008b. {A statistical mechanics model for free-for-all
  airplane passenger boarding}. American Journal of Physics 76, 1114--1119.

\bibitem[{van~den Briel et~al.(2005)van~den Briel, Villalobos, Hogg, Lindemann,
  and Mule}]{VandenBriel:2005}
van~den Briel, M. H.~L., Villalobos, J.~R., Hogg, G.~L., Lindemann, T., Mule,
  A.~V., 2005. America west airlines develops efficient boarding strategies.
  INTERFACES 35~(3), 191--201.

\end{thebibliography}







\end{document}